\documentclass[aip,apl,reprint,a4paper,superscriptaddress,floatfix,amsmath,amssymb,amsfonts,noshowpacs,longbibliography]{revtex4-2}

\usepackage{newtxtext,newtxmath}
\usepackage[utf8]{inputenx}
\input{ix-utf8enc.dfu}
\usepackage{graphicx}
\usepackage[dvipsnames]{xcolor}
\usepackage{soul} 
\usepackage{xspace}
\usepackage{siunitx}

\usepackage[
,textwidth=17.5cm
,textheight=23.7cm
,verbose
,pdftex
]{geometry}
\usepackage{xspace}

\usepackage[pdftex]{hyperref}
\hypersetup{
	pdftitle={Lattice parameters of (Sc,Al)N layers grown on GaN(0001) by plasma-assisted molecular beam epitaxy},
	colorlinks,
	citecolor=blue,
	linkcolor=blue,
	urlcolor=blue
}

\newcommand{\celsius}{$^{\circ}$C\xspace}
\newcommand{\scaln}{Sc$_x$Al$_{1-x}$N\xspace}

\begin{document}

\title{Lattice parameters of Sc$_{\boldsymbol{\mathsf{x}}}$Al$_{\boldsymbol{\mathsf{1-x}}}$N layers grown on GaN(0001) by plasma-assisted molecular beam epitaxy}

\author{Duc V. Dinh}
\email[Electronic email: ]{dinh@pdi-berlin.de}
\author{Jonas Lähnemann}
\author{Lutz Geelhaar}
\author{Oliver Brandt}
\affiliation{Paul-Drude-Institut für Festkörperelektronik, Leibniz-Institut im Forschungsverbund Berlin e.V., Hausvogteiplatz 5--7, 10117 Berlin, Germany.}


\begin{abstract}
An accurate knowledge of the lattice parameters of the new nitride \scaln is essential for understanding the elastic and piezoelectric properties of this compound as well as for the ability to engineer its strain state in heterostructures. Using high-resolution x-ray diffractometry, we determine the lattice parameters of 100-nm-thick undoped \scaln layers grown on GaN(0001) templates by plasma-assisted molecular beam epitaxy. The Sc content $x$ of the layers is measured independently by both x-ray photoelectron spectroscopy and energy-dispersive x-ray spectroscopy and ranges from 0 to 0.25. The in-plane lattice parameter of the layers linearly increases with increasing $x$, while their out-of-plane lattice parameter remains constant. Layers with $x\approx0.09$ are found to be lattice matched to GaN, resulting in a smooth surface and a structural perfection equivalent to that of the GaN underlayer. In addition, a two-dimensional electron gas is induced at the \scaln/GaN heterointerface, with the highest sheet electron density and mobility observed for lattice-matched conditions. 

\end{abstract}

                  
\maketitle


In the last decade, \scaln ternary alloys have gained considerable attention for their outstanding piezoelectric properties,\cite{Akiyama2009Feb,Matloub2013Apr} making them attractive for applications in surface acoustic wave (SAW) devices \cite{Hashimoto2013Mar,Zhang2015Sep,Kobayashi2021Jan} as well as for heterostructure field-effect transistors.\cite{Hardy2017Apr, Frei2019May, Ligl2020May} Specifically, the longitudinal piezoelectric strain coefficient $d_{33}$ along the [0001] axis has been found to increase with the Sc content $x$, as long as the wurtzite structure of \scaln is maintained. For example, $d_{33}$ of Sc$_{0.43}$Al$_{0.57}$N has been found to be four times higher than that of pure AlN.\cite{Akiyama2009Feb} This high value of $d_{33}$ results in an increase of the electromechanical coupling factor, enabling a more efficient conversion of electrical to mechanical energy and vice versa. Furthermore, the large polarization discontinuity between \scaln and GaN results in unprecedentedly high electron densities and correspondingly low sheet resistances of \scaln/GaN heterostructure field-effect transistors.\cite{Hardy2017Apr, Frei2019May, Ligl2020May, Manz2021Jan, Wang2021Aug} Finally, very recent work has shown that \scaln is even a ferroelectric material and thus may have applications in this evolving area.\cite{Fichtner2019Mar, Wang2021May, Wang2022Jul, Wang2023Jan}

In the past, \scaln films have been mostly fabricated by magnetron sputtering.\cite{Akiyama2009Feb, Matloub2013Apr, Hoglund2010Jun, Zukauskaite2012May, Deng2013Mar, Saha2015Feb, Knisely2019Apr, Osterlund2021Mar} These films are invariably characterized by a comparatively low crystallinity for thicknesses exceeding 1\,\textmu m.\cite{Akiyama2009Feb, Zukauskaite2012May, Knisely2019Apr, Osterlund2021Mar} Recently, \scaln layers with higher crystallinity have been produced on GaN templates by both plasma-assisted molecular beam epitaxy (PAMBE) \cite{Hardy2017Apr, Wang2020Apr, Dargis2020Apr, Hardy2020May, Casamento2020Sep} and metal-organic vapor phase epitaxy (MOVPE).\cite{Ligl2020May, Leone2020Jan, Manz2021Jan} However, despite the improved structural properties of these layers, the existing studies do not even agree on one of the most basic properties of a crystalline substance, namely, the lattice parameters. In particular, the individual studies report widely differing dependencies of the in-plane lattice parameter \textit{a} on the Sc content $x$, and a large scatter of the out-of-plane lattice parameter \textit{c}.\cite{Akiyama2009Feb, Matloub2013Apr, Hardy2017Apr, Hoglund2010Jun, Zukauskaite2012May, Deng2013Mar, Knisely2019Apr, Osterlund2021Mar,Wang2020Apr, Dargis2020Apr, Hardy2020May, Casamento2020Sep, Leone2020Jan} 
The \textit{c/a} ratio of \scaln is predicted to basically determine its piezoelectric coefficients,\cite{Akiyama2009Feb, Matloub2013Apr, Momida2018Feb} i.\,e., the understanding of the elastic and piezoelectric properties of this material rests on the determination of the lattice parameters of \scaln. Moreover, an accurate knowledge of an alloy's lattice parameters is a basic prerequisite for using this material in applications demanding control over composition and strain, such as \scaln/GaN heterostructure field-effect transistors.

In this letter, we determine the lattice parameters of 100-nm-thick \scaln ($0\! \leq \! x \!<\! 0.25$) layers grown on GaN by PAMBE. For a reliable determination of the Sc content $x$ of the layers, we employ two independent analytical techniques, namely, x-ray photoelectron spectroscopy (XPS) and energy-dispersive x-ray spectroscopy (EDX). The lattice parameters of the layers are measured using triple-axis high-resolution x-ray diffractometry (HRXRD). The comparatively thick layers facilitate the discrimination between a pseudomorphic strain state and actual lattice matching, as well as the detection of asymmetric XRD reflections allowing an accurate determination of both the out-of-plane and in-plane lattice parameters.

\scaln layers are grown by PAMBE on quarters of unintentionally doped 4-µm-thick GaN/Al$_2$O$_3$(0001) 2-inch wafers. Before being loaded into the ultrahigh vacuum environment, the GaN templates are etched in HCl solution to remove the surface oxide as well as surface contaminants, and then rinsed with de-ionized water and finally blown dry with a nitrogen gun. Afterwards, the templates are outgassed for 2 hours at 500\,°C in a load-lock chamber attached to the MBE system. The MBE growth chamber is equipped with high-temperature effusion cells to provide the group III metals (99.9999\,\% pure Al and 99.995\,\% pure Sc). A Veeco UNI-Bulb radio-frequency plasma source is used for the supply of active nitrogen (N$^*$). The N$^*$ flux is calculated from the thickness of a GaN layer grown under Ga-rich conditions and thus with a growth rate limited by the N$^*$ flux. Prior to \scaln growth, a 100-nm-thick GaN buffer layer is grown at 700\,°C under Ga-bilayer conditions. For the subsequent \scaln layer, the growth temperature as measured by a thermocouple is set to 700--725\,°C. To vary $x$, the layers are grown with different Sc fluxes while keeping the fluxes of Al and N$^*$ constant, resulting in a growth rate of 2.5--3.0\,nm/min. The corresponding III/N$^*$ ratio is varied from 0.45 to 0.6. To ensure compositional uniformity across the wafers, the samples are rotating during growth with a speed of three revolutions per minute. The structural properties of the layers are characterized using an HRXRD system (Philips Panalytical X'Pert PRO MRD) equipped with a two-bounce hybrid monochromator Ge(220) for the CuK$_{\alpha1}$ source ($\uplambda=\SI{1.540598}{\Angstrom}$). The thickness of the layers of 100--120\,nm is obtained from x-ray thickness interference fringes (see Fig.\,S1 in the supplementary material). The lattice parameters of the layers are calculated from the angular positions of the symmetric 0002 and asymmetric 10\=15 reflections in $2\theta$-$\omega$ scans and reciprocal space maps (RSMs), respectively, both measured with analyzer. X-ray rocking curves (XRCs) across the 0002 and 10\=12 reflections are measured with an open detector without any receiving slit. The composition of the layers is independently measured by XPS (Scienta Omicron) and EDX (EDAX Octane Elect Super) using the thin-film analysis software BadgerFilm.\cite{Moy2021Apr} The latter was shown to yield layer thicknesses and compositions in excellent agreement with Rutherford backscattering spectrometry for multilayer structures. Note that XPS returns the composition of the top 10\,nm of the layer, while EDX averages over the whole thickness of the layer. The values obtained from these techniques are close to each other with an average mean absolute deviation of $\pm 0.6$\% (see Fig.\,S2 in the supplementary material). We take the mean of these values to represent the actual Sc content $x$. The surface morphology of the layers is imaged by atomic force microscopy in contact mode (Dimension Edge, Bruker). To investigate the electrical properties of the layers, Hall effect measurements in the van der Pauw configuration are performed at room temperature.

\begin{figure}[t]
	\centering
	\includegraphics[width=\columnwidth]{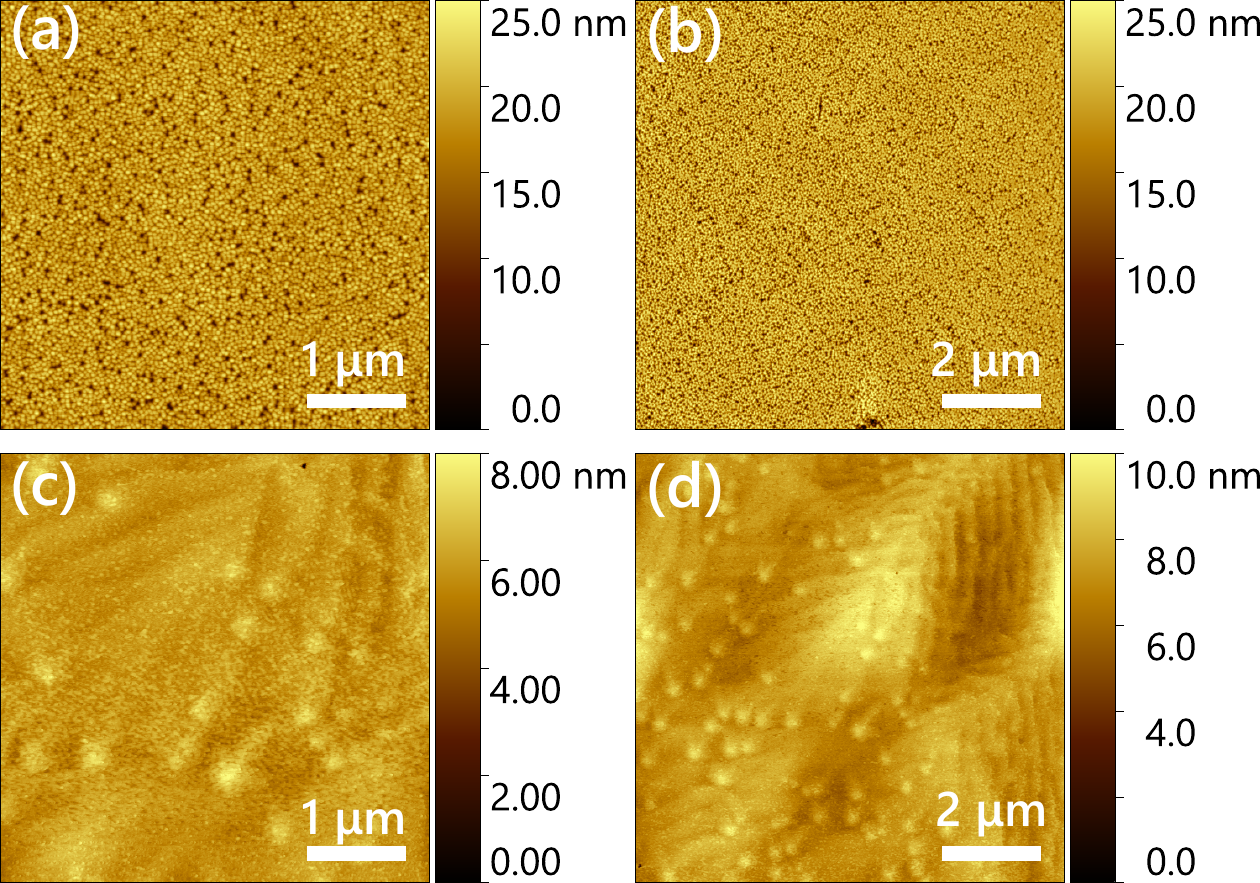}
	\caption{(a,\,c) $5\times5$\,$\mu$m$^{2}$ and (b,\,d) $10\times10$\,$\mu$m$^{2}$ atomic force topographs of 105-nm-thick Sc$_{0.09}$Al$_{0.91}$N layers grown on GaN at 725\,°C (top row) and 700\,°C (bottom row), respectively. Root-mean square roughness values of (a) 2.8\,nm, (b) 3.3\,nm, (c) 0.5\,nm and (d) 0.9\,nm are obtained from these topographs.}
	\label{fig:AFM}
\end{figure}

For pure AlN, the optimum growth conditions in PAMBE for obtaining smooth layers with a high structural integrity consist of an Al/N ratio slightly higher than one and a temperature above 820\,\celsius resulting in Al-stable conditions with a mono- to bilayer Al adlayer on the growth front.\cite{Ive2005Jan} Lower temperatures suffice for (Al,Ga)N since the more volatile Ga adatoms can be employed for forming a stable metal adlayer. For (Sc,Al)N, however, temperatures at least as high as for pure AlN would have to be used for implementing metal-stable growth conditions, as only Al adatoms have a finite desorption rate at temperatures achievable in PAMBE, while the desorption of Sc adatoms is negligible. 

In contrast to these considerations, we obtain smooth \scaln layers only at much lower temperatures, which require growth with a III/N$^*$ ratio below one. Figure~\ref{fig:AFM} shows atomic force topographs of 105-nm-thick Sc$_{0.09}$Al$_{0.91}$N layers grown on GaN at 725\,\celsius (a,\,b) and 700\,\celsius (c,\,d), respectively. The layers' morphology is seen to change from a rough surface characterized by a dense arrangement of faceted hillocks (a,\,b) intersected by pits with a density of about $10^9$\,cm$^{-2}$ to a smooth surface with monatomic steps by decreasing the substrate temperature by only 25\,\celsius. For the lower temperature, the root-mean-square roughness stays below 1\,nm on an area of $10 \times 10$\,\textmu m$^2$, which is a remarkably low value considering that the layers were grown with a considerable N excess (III/N$^* \approx 0.5$). For all layers presented in the following (for their topographs, see Fig.\,S3 in the supplementary material), we have therefore chosen a growth temperature of 700\,\celsius.

\begin{figure}[t]
	\includegraphics[width=\columnwidth]{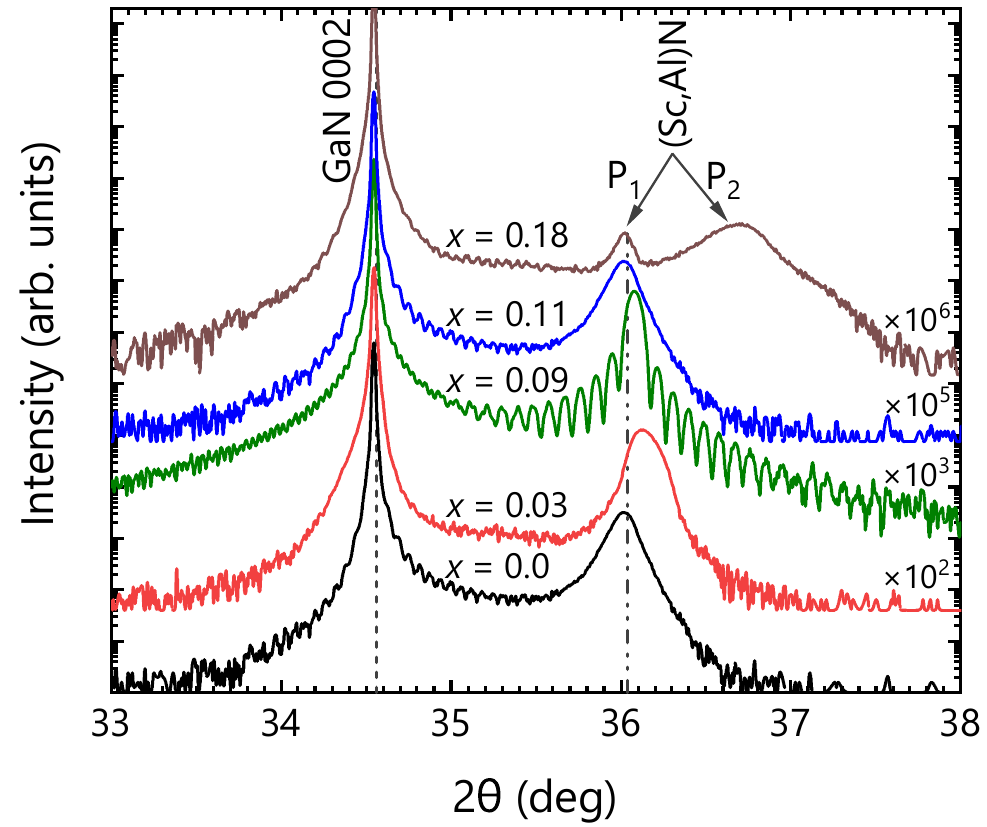}
	\caption{Symmetric $2\theta$-$\omega$ XRD scans of \scaln layers grown on GaN(0001) templates with different Sc content $x$. The individual scans are vertically shifted for clarity as indicated in the figure. The dashed and dash-dotted lines indicate the 2$\theta$ position of the 0002 reflection of relaxed GaN and AlN, respectively.}
	\label{fig:XRD}
\end{figure}

Figure~\ref{fig:XRD} shows symmetric $2\theta$-$\omega$ XRD scans of Sc$_x$Al$_{1-x}$N layers grown on GaN(0001) templates with different Sc content $x$. All the layers exhibit a reflection (P$_1$) at about 36.0--36.1° that is close to the AlN\,0002 reflection. For the layers with $x>0.15$, an additional broad reflection (P$_2$) appears at about 36.5--37.0°, indicating a phase separation. Reflections pertinent to Al$_3$Sc\,111 have not been observed for any of the samples (see Fig.\,S4 in the supplementary material),\cite{Elliott1981Sep} confirming the benefit of N$^*$-rich conditions for avoiding the formation of this intermetallic phase in the PAMBE of \scaln.\cite{Frei2019May, Hardy2017Apr, Hardy2020May, Casamento2020Sep, Schuster1985Jul}

\begin{figure}[t]
	\centering
	\includegraphics[width=\columnwidth]{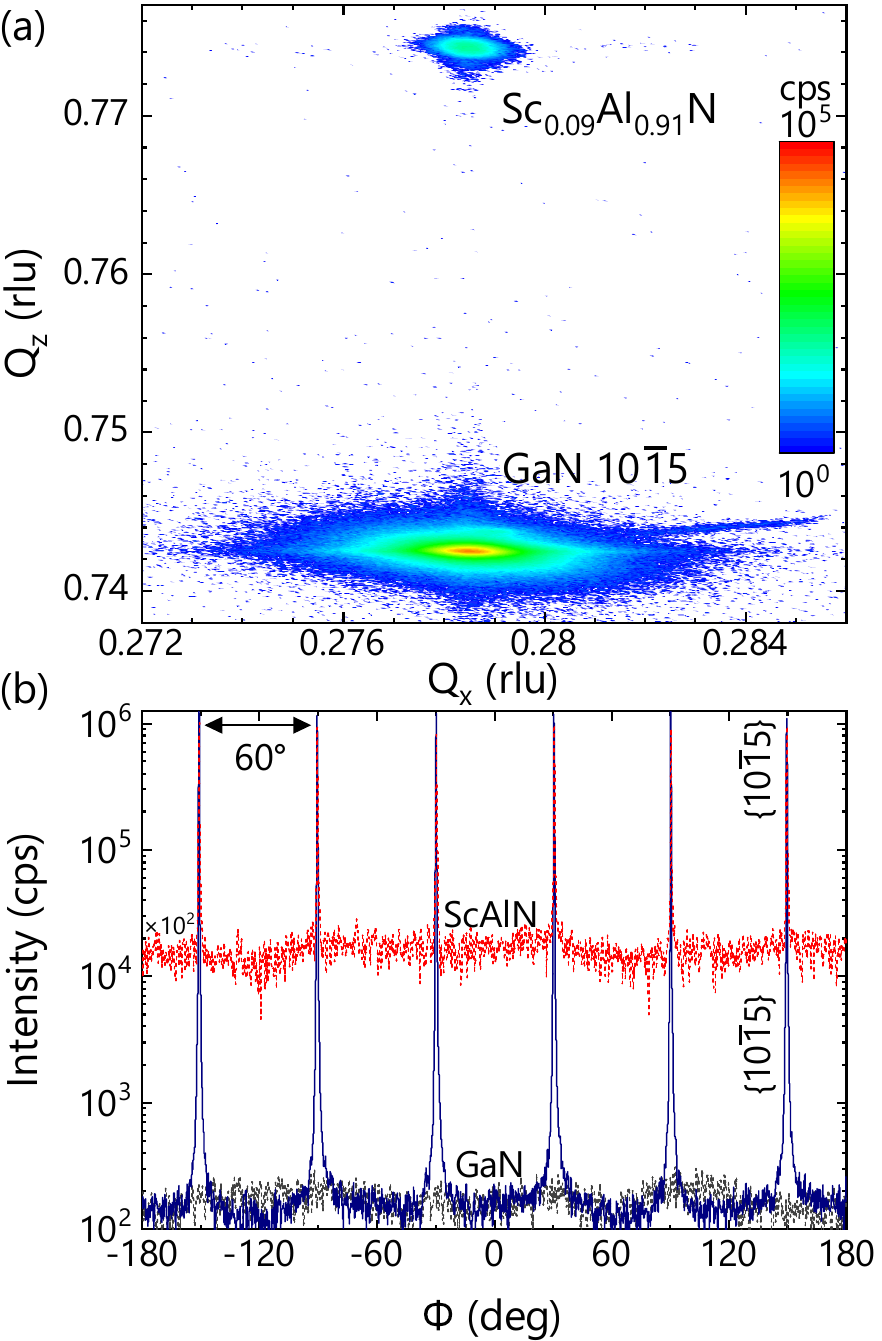}
	\caption{(a) RSM in the vicinity of the asymmetric GaN\,10\=15 reflection of the 105-nm-thick Sc$_{0.09}$Al$_{0.91}$N/GaN structure. The RSM is displayed in reciprocal lattice units (rlu). (b) Azimuthal $\phi$ scans of the Sc$_{0.09}$Al$_{0.91}$N and GaN\,10\=15 reflections performed with an open detector without receiving slit.} 
	\label{fig:PhiRSM}
\end{figure}

To determine the lattice parameters of the \scaln layers, we record RSMs around the GaN\,10\=15 reflection. Figure~\ref{fig:PhiRSM}(a) shows an exemplary RSM of the 105-nm-thick Sc$_{0.09}$Al$_{0.91}$N/GaN layer (RSMs of other samples are presented in Fig.\,S5 in the supplementary material). Besides the GaN\,10\=15 peak, only a single reflection related to the Sc$_{0.09}$Al$_{0.91}$N layer is observed. Assuming that this layer has a wurtzite crystal structure with (0001) orientation, the in-plane and out-of-plane lattice parameters can be estimated from the Q$_x$ and Q$_z$ axes of the RSM by the following equation:  
\def\A{
\begin{pmatrix}
    a \\
    c
\end{pmatrix}}
\def\B{
\begin{pmatrix}
    1/\sqrt{3} 	& 0 \\
    0 					& 5/2
\end{pmatrix}}
\def\C{
\begin{pmatrix}
    1/Q_x \\
    1/Q_z
\end{pmatrix}}
\begin{equation}
\A = \uplambda \B \C.
\label{eq:matrix}
\end{equation}
Based on the lattice parameters determined using Eq.\,\ref{eq:matrix}, we perform azimuthal scans of the Sc$_{0.09}$Al$_{0.91}$N\,10\=15 (2$\theta$\,=\,110.5°, $\omega$\,=\,55.3°, $\phi$\,=\,0°, $\psi$\,=\,20°) and GaN\,10\=15 (2$\theta$\,=\,105.0°, $\omega$\,=\,52.5°, $\phi$\,=\,0°, $\psi$\,=\,20.6°) reflections in skew geometry.
As shown in Fig.~\ref{fig:PhiRSM}(b), the scans of both the GaN and Sc$_{0.09}$Al$_{0.91}$N layers exhibit six equidistant maxima separated by 60° and appearing at the same angular positions. This result reveals a shared sixfold symmetry of the wurtzite Sc$_{0.09}$Al$_{0.91}$N/GaN heterostructure and a strict in-plane orientation-relationship between the substrate and the layer, thus allowing us to identify the reflection of the Sc$_{0.09}$Al$_{0.91}$N layer in the RSM (Fig.~\ref{fig:PhiRSM}(a)) as the 10\=15 reflection. Having the same value of $Q_x$, the Sc$_{0.09}$Al$_{0.91}$N and the GaN layers therefore share the same in-plane lattice parameter, meaning that the Sc$_{0.09}$Al$_{0.91}$N layer is either pseudomorphic or lattice-matched to the GaN template. This finding is further confirmed by the fact that a 350-nm-thick, fully relaxed Sc$_{0.09}$Al$_{0.91}$N on AlN/Al$_2$O$_3$ exhibits exactly the same lattice constant (see Fig.\,S6 in the supplementary material).

For \scaln layers with $x\!\!>\!\!0.15$, their 10\=15 reflection is too weak and broad (and even undetectable for the Sc$_{0.24}$Al$_{0.76}$N layer) to be located with certainty, even when performing the scans with a 1/4° receiving slit instead of an analyzer. This deterioration of the layers' crystallinity coincides with the occurrence of the additional reflection P$_2$ at 36.5--37.0° (Fig.~\ref{fig:XRD}), presumably reflecting a beginning phase separation. Note that the reflection originating from \scaln\,0002 (P$_1$ in Fig.~\ref{fig:XRD}) continues to be detected, i.\,e., the phase separation does not occur abruptly and uniformly, but we are dealing with a coexistence of wurtzite and rocksalt \scaln. This conclusion is in agreement with the findings of \citet{Constantin2004Nov} for Sc$_y$Ga$_{1-y}$N layers grown by PAMBE, who obtained layers with pure wurtzite crystal structure for $y\! \leq \! 0.17$, followed by a gradual transition from the wurtzite to the rocksalt structure for $0.17\!<\!y\leq\!0.54$, and finally the rocksalt structure for $y\! > \! 0.54$. For sputtered \scaln, the transition from the wurtzite to the rocksalt structure was reported to take place for $x>0.20\pm0.03$,\cite{Deng2013Mar, Hoglund2010Jun, Saha2015Feb} in agreement with the critical value observed in PAMBE. Similarly, wurtzite inclusions in sputtered cubic \scaln have been observed at $x = 0.1$ by selected area electron diffraction.\cite{Hoglund2009Jun} Accordingly, we identify reflection P$_2$ as stemming from rocksalt \scaln. Linearly interpolating between the measured lattice parameters of rocksalt AlN\cite{Madan1997Mar} and ScN,\cite{Al-Brithen2000Oct} the angular position of P$_2$ corresponds to a Sc content of 0.3--0.4.

\begin{figure}[t]
	\centering
	\includegraphics[width=\linewidth]{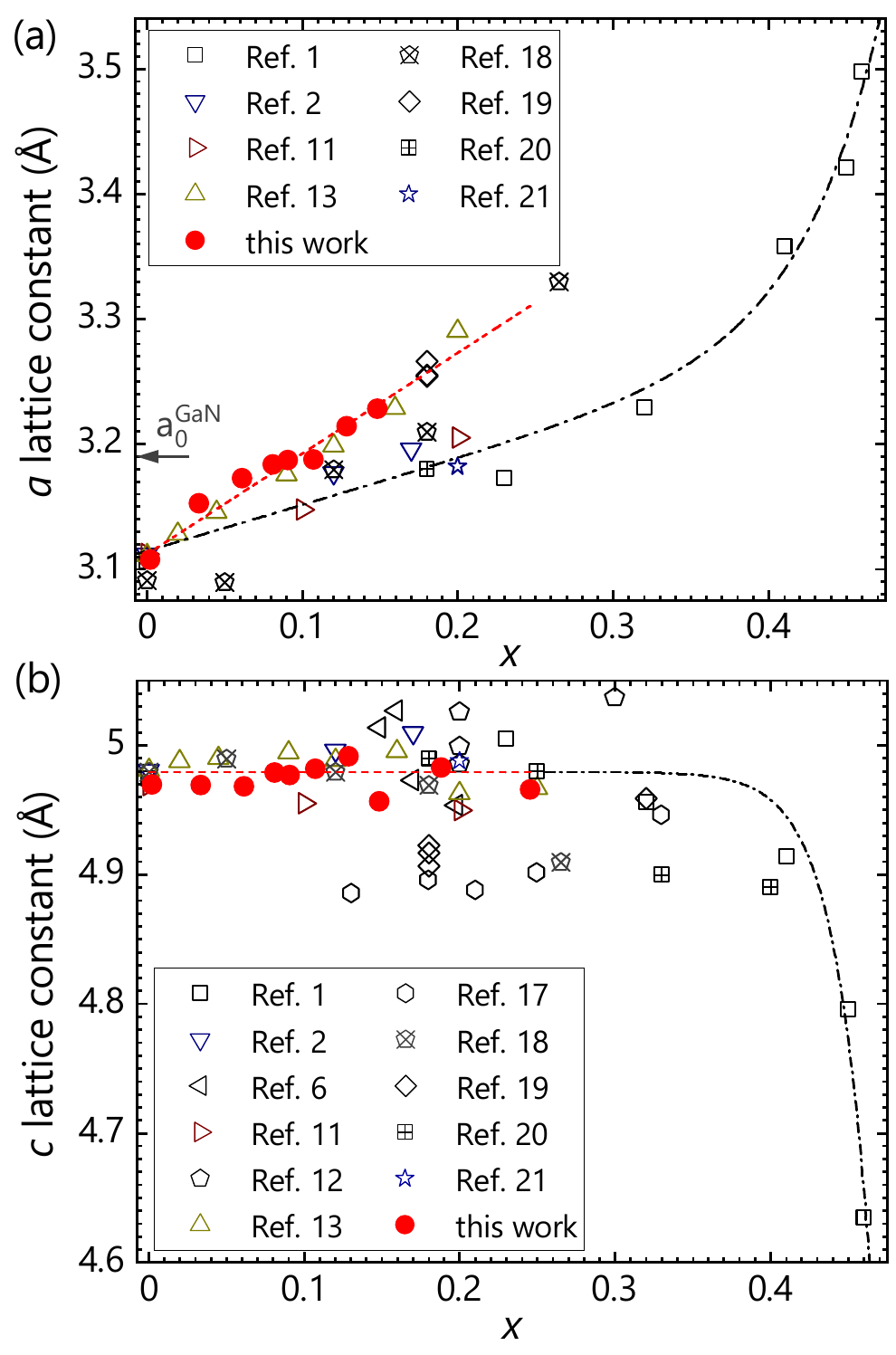}
	\caption{(a) In-plane and (b) out-of-plane lattice parameters $a$ and $c$, respectively, as a function of $x$ for Sc$_x$Al$_{1-x}$N layers on GaN (solid circles). Various values from the literature are included for comparison with symbols as indicated in the legend. The lines in both figures are guides to the eye. The arrow in (a) indicates the relaxed in-plane lattice parameter (\textit{a}$^{\text{GaN}}_{0}$) of GaN.}
	\label{fig:lattice}
\end{figure}

Figure \ref{fig:lattice} summarizes the in-plane ($a$) and out-of-plane ($c$) lattice parameters of \scaln layers measured in the present work and in previous studies of other research groups. The former is shown in Fig.~\ref{fig:lattice}(a) and is seen to linearly increase with $x$ following the dependence $a(x) = 3.112 + 0.85 x$, while the latter shown in Fig.~\ref{fig:lattice}(b) is essentially  constant. This anomalous behavior was first observed by \citet{Constantin2004Nov} for Sc$_y$Ga$_{1-y}$N and later also for \scaln,\cite{Akiyama2009Feb, Hoglund2010Jun, Deng2013Mar, Matloub2013Apr,Hoglund2009Jun} and has been attributed to the distortion of the wurtzite structure due to the existence of a metastable hexagonal phase of ScN.\cite{Constantin2004Nov,Deng2013Mar}

The $a(x)$ dependence observed in our work is in excellent agreement with the results obtained in Refs.~\citenum{Deng2013Mar} and \citenum{Hardy2020May}, but differs in slope for $x \! < \! 0.25$ by a factor of two from the results reported in Refs.~\citenum{Akiyama2009Feb,Matloub2013Apr,Hoglund2010Jun,Casamento2020Sep,Leone2020Jan}. As a result, lattice matching is expected for the former results at $x \! \approx \! 0.09$ and for the latter at  $x \! \approx \! 0.18$. The reason for this discrepancy is likely not related to the deposition method, or the crystallinity of the film, since both sets of data include results stemming from layers prepared by sputtering or epitaxial techniques. \citet{Casamento2020Sep} concluded that lattice-matching to GaN occurs for $x \! = \! 0.18$ on the basis of equal in-plane lattice parameters for GaN and a 28-nm-thick Sc$_{0.18}$Al$_{0.82}$N layer. However, such a thin layer may grow pseudomorphically (i.\,e., coherently strained) on GaN with the same in-plane constant even with a composition quite far from lattice matching. Specifically, the lattice mismatch between Sc$_{0.18}$Al$_{0.82}$N and GaN amounts to about 2\% according to our data, and this mismatch may still be accommodated coherently at a thickness of 28\,nm. For the layer thickness of about 100\,nm  considered in the present work, the tolerance for deviations from lattice matching is much lower, allowing us to determine the composition corresponding to lattice matching more accurately. Thicker (350\,nm) films with this composition on a highly mismatched AlN buffer layer are indeed found to have the same in-plane and out-of-plane lattice parameters (see Fig.\,S6 in the supplementary material).   

Regarding the $c$ lattice parameter, our data are in good agreement with the values reported by Ref.~\citenum{Deng2013Mar} over the whole compositional range up to $x\!=\!0.25$, and with isolated values from Refs.~\citenum{Matloub2013Apr,Dargis2020Apr,Leone2020Jan,Casamento2020Sep}. The values reported in the literature widely scatter for $x\!>\!0.15$, deviating by as much as $\pm 0.07$\,nm or $\pm 1.5$\%, which is a very large deviation for a lattice parameter. Presumably, the determination of the lattice parameters is hampered in this range of Sc contents by the transition from the wurtzite to the rocksalt structure as discussed above.\cite{Constantin2004Nov, Matloub2013Apr}  

\begin{figure}[t]
	\centering
	\includegraphics[width=\columnwidth]{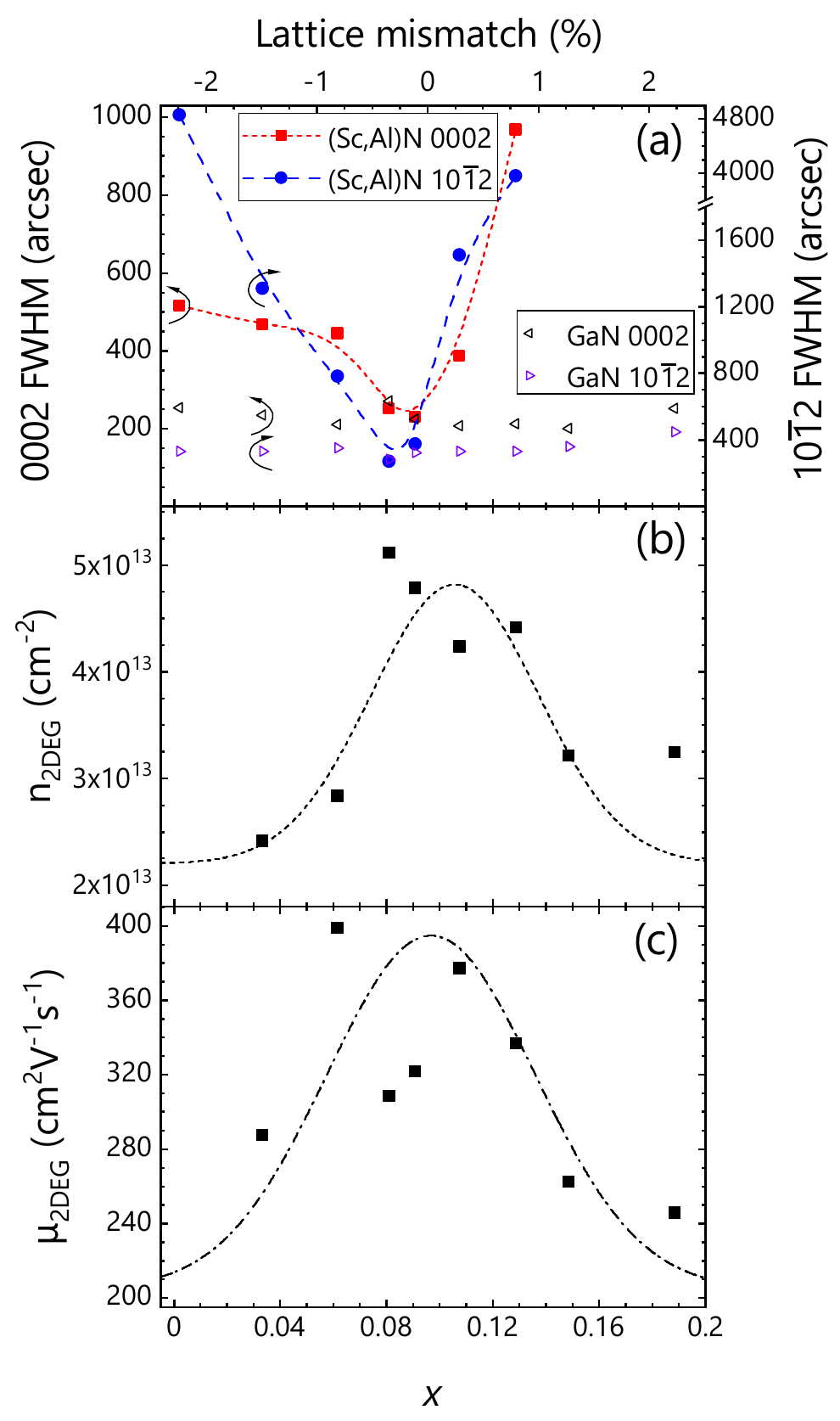}
	\caption{Dependence of (a) the values of the FWHM of 0002 and 10\=12 XRCs, (b) the Hall sheet carrier density and (c) the electron mobility of \scaln/GaN layers on the Sc content $x$. The lines are a guide to the eye.}
	\label{fig:XRC_Hall}
\end{figure}

Finally, we examine the impact of lattice mismatch on the structural perfection of the \scaln layers and the \scaln/GaN heterointerface. Figure~\ref{fig:XRC_Hall}(a) shows values of the FWHM of the 0002 and 10\=12 XRCs of the samples under investigation. Both reach minimum values for lattice-matched conditions at $x\!=\!0.09$ and for a slightly tensile strain of $-0.5$\% at $x\!=\!0.05$. In both cases, the values for the FWHM are comparable to those of the GaN templates, namely, 240 and 270--370\,arcsec for the 0002 and 10\=12 XRCs, respectively. For larger tensile and compressive strains, the values of the FWHM are seen to increase drastically, particularly so for the 10\=12 reflection. Note that XRCs of asymmetric reflections have not been performed in any previous work.

The polarization discontinuity between \scaln and GaN is expected to induce a two-dimensional electron gas (2DEG) at their interface, which we probe here by room-temperature Hall-effect measurements. However, the GaN templates we have used as substrates are not semi-insulating, but unintentionally doped, and electrical conduction will thus take place in them as well. Hence, we analyze the Hall-effect results by a two-layer model of transport to extract the actual electron sheet density $n_\text{2DEG}$ and mobility $\mu_\text{2DEG}$ in the 2DEG.\cite{Petritz1958Jun, Look1997Jun, Arnaudov2003Jan} Solving the equations for $n_\text{2DEG}$ and $\mu_\text{2DEG}$ while assuming the contribution from the GaN layer to be constant and equal to the independently measured values from a bare GaN template (electron sheet density of $2\times10^{13}$\,cm$^{-2}$ and a mobility of 200\,cm$^{2}$V$^{-1}$s$^{-1}$) yields the values displayed in Figs.~\ref{fig:XRC_Hall}(b) and \ref{fig:XRC_Hall}(c). Clearly, both the sheet density and the mobility of the 2DEG have a maximum close to lattice matching and sharply decline away from this maximum for both tensile and compressive strain. The highest electron sheet density and mobility of $4.5\times10^{13}$\,cm$^{-2}$ and 400\,cm$^{2}$V$^{-1}$s$^{-1}$ are comparable with those reported for 10--23\,nm-thick \scaln/GaN layers grown by MBE\cite{Hardy2017Apr, Frei2019May, Wang2021Aug} and MOVPE.\cite{Ligl2020May} For the layers investigated here, the reduction of both $n_\text{2DEG}$ and $\mu_\text{2DEG}$ for $x \lessgtr 0.1$ is attributed to the formation of cracks (for tensile strain---see Fig.\,S3 in the supplementary material) and dislocations at the \scaln/GaN interface,\cite{Jena2000Mar} as well as interface roughness and alloy scattering for higher Sc content.\cite{Rajan2006Jan} 

To summarize and conclude, we have presented an accurate determination of the lattice parameters of 100-nm-thick \scaln $(0 \! \leq \! x \! \leq \! 0.25)$ layers grown on GaN(0001) templates by plasma-assisted molecular beam epitaxy. The comparatively thick layers allow us to more easily discriminate between a pseudomorphic strain state and actual lattice matching. Furthermore, the Sc content was determined by two independent techniques to obtain reliable values of $a(x)$ and $c(x)$. Lattice matched Sc$_{0.09}$Al$_{0.91}$N layers have been found to exhibit excellent structural and morphological properties, as well as sustaining a two-dimensional electron gas with high sheet density and mobility.

\small{See the supplementary material for (1) experimental and simulated 2$\theta$-$\omega$ XRD scans of the Sc$_{0.09}$Al$_{0.91}$N layer; (2) the Sc content independently estimated by XPS and EDX; (3) AFM topographs of the Sc$_x$Al$_{1-x}$N layers; (4) 2$\theta$-$\omega$ XRD scans of \scaln layers recorded over a wide angular range with open detector; (5) asymmetric RSMs of 100-nm-thick (Sc,Al)N layers grown on GaN; (6) asymmetric RSM of a 350-nm-thick Sc$_{0.09}$Al$_{0.91}$N layer grown on an AlN template.

The data that supports the findings of this study are available within the article and its supplementary material.

We thank Oliver Bierwagen for a critical reading of the manuscript and Carsten Stemmler for expert technical assistance with the MBE system. 
}

\bibliography{MyBIB}

\end{document}




\title{Lattice parameters of Sc$_{\boldsymbol{\mathsf{x}}}$Al$_{\boldsymbol{\mathsf{1-x}}}$N layers grown on GaN(0001) by plasma-assisted molecular beam epitaxy}

\author{Duc V. Dinh}
\email[Electronic email: ]{dinh@pdi-berlin.de}
\author{Jonas Lähnemann}
\author{Lutz Geelhaar}
\author{Oliver Brandt}
\affiliation{Paul-Drude-Institut für Festkörperelektronik, Leibniz-Institut im Forschungsverbund Berlin e.V., Hausvogteiplatz 5--7, 10117 Berlin, Germany}

\maketitle

\begin{figure}
	\centering
	\includegraphics[width=0.75\linewidth]{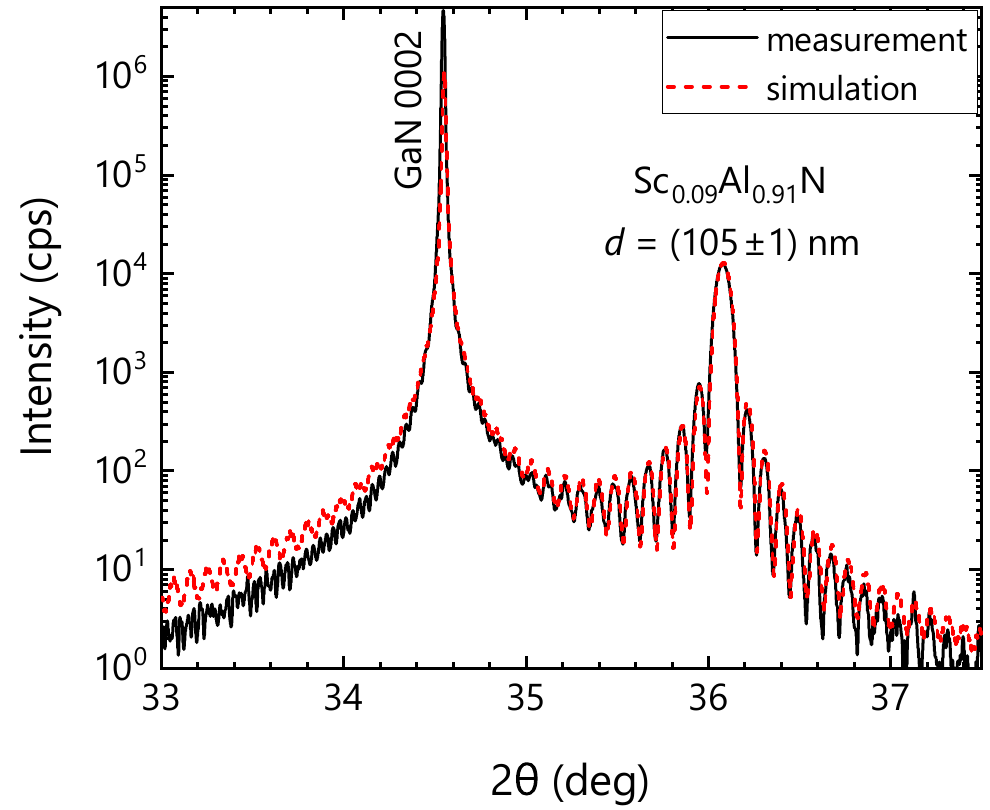}
	\caption{Experimental and simulated 2$\theta-\omega$ XRD scans of the 105-nm-thick Sc$_{0.09}$Al$_{0.91}$N layer.}
\end{figure}

\clearpage
\bibliography{MyBIB}

\begin{figure}
	\centering
	\includegraphics[width=0.75\linewidth]{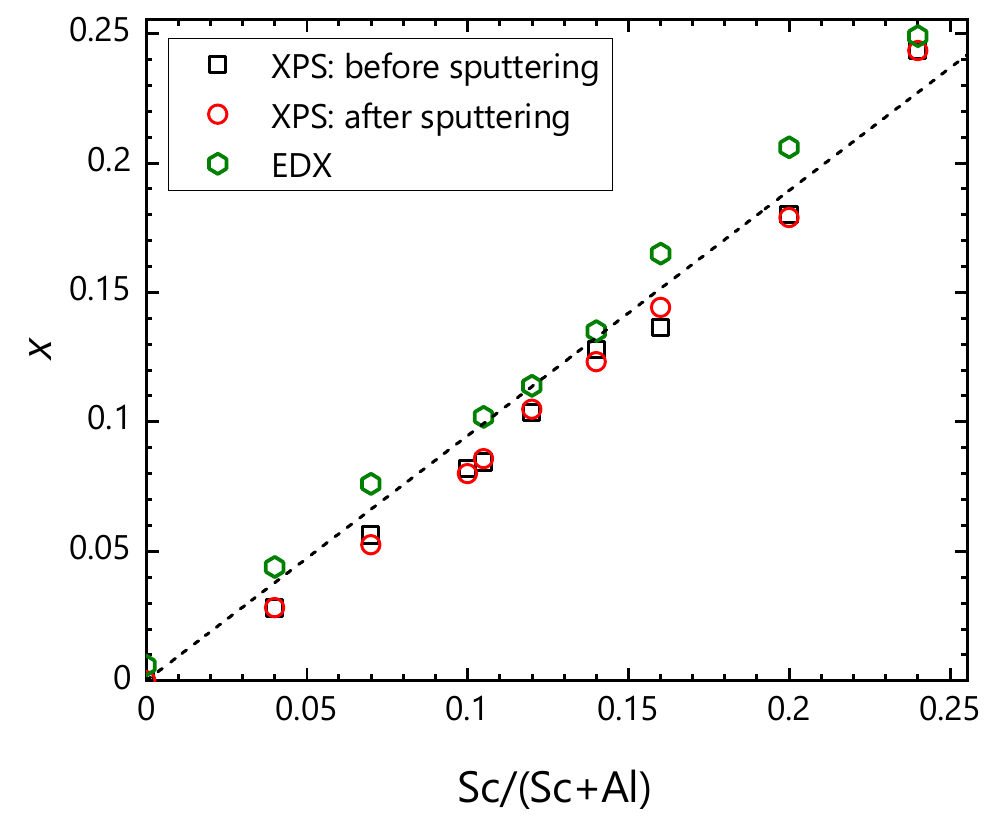}
	\caption{Sc content \textit{x} of the Sc$_x$Al$_{1-x}$N layers under investigation determined by both x-ray photoelectron spectroscopy (XPS) and energy-dispersive x-ray spectroscopy (EDX) as a function of the flux ratio Sc/(Sc+Al). The line is a guide to the eye. XPS was measured before and after removing surface contaminants by Ar$^{+}$ sputtering. For EDX, we conducted a series of experiments with varied acceleration voltage, accompanied by reference measurements on binary ScN and AlN layers, and analyzed the spectra using the thin-film analysis software BadgerFilm.~\cite{Moy2021Apr} The values determined by these two methods agree with each other on average within $\pm 0.6$\%.\\\\\\\\\\}
\end{figure}

\begin{figure}
	\centering
	\includegraphics[width=0.75\linewidth]{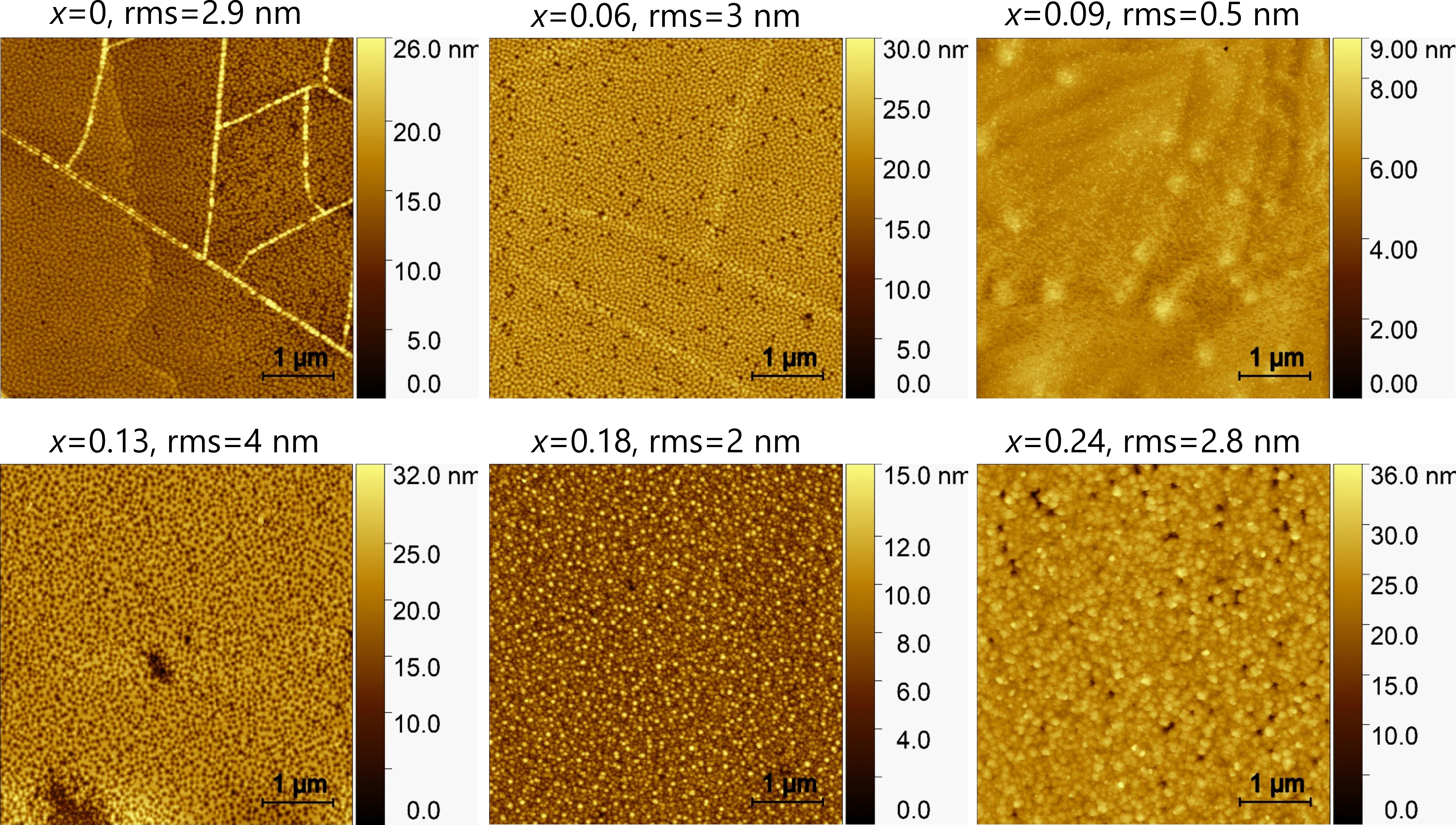}
	\caption{$5\times5$\,$\mu$m$^{2}$ atomic force topographs of 100-nm-thick Sc$_{x}$Al$_{1-x}$N layers grown on GaN templates. Root-mean square roughness (rms) values obtained from these topographs are shown for comparison. Cracks can be seen on the topographs of samples grown with $x=0$ and $x=0.06$.}
\end{figure}

\begin{figure}
	\centering
	\includegraphics[width=0.75\linewidth]{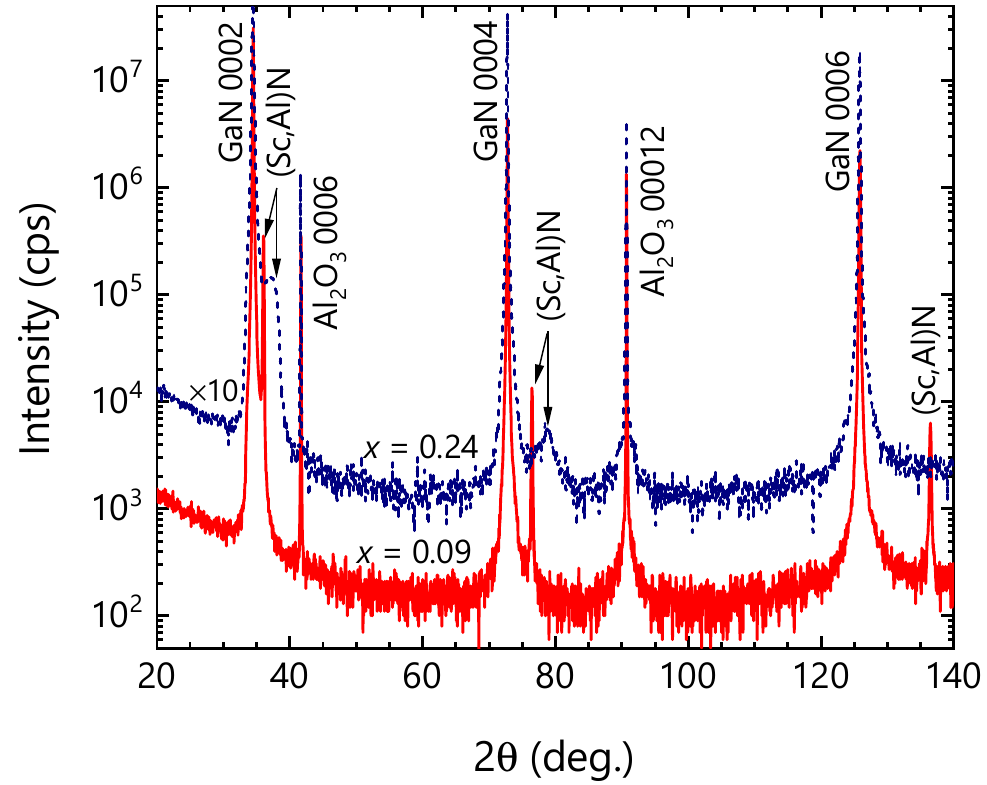}
	\caption{Symmetric 2$\theta-\omega$ XRD scans of the Sc$_{0.09}$Al$_{0.91}$N and Sc$_{0.245}$Al$_{0.755}$N layers measured with open detector without any receiving slit. No secondary reflections related to the intermetallic compound Al$_{y}$Sc (111) (\textit{y}\,=\,1,2,3) are observed.}
\end{figure}

\begin{figure}
	\centering
	\includegraphics[width=\linewidth]{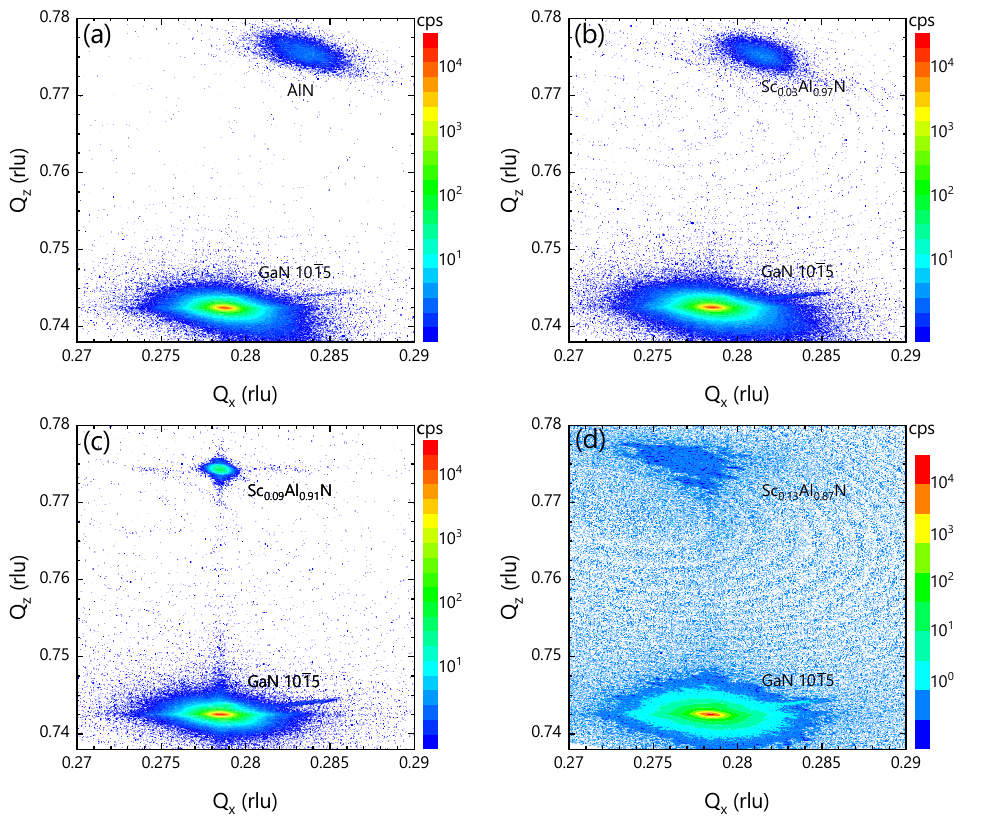}
	\caption{RSM in the vicinity of the asymmetric GaN\,10\=15 reflection of (a) AlN/GaN, (b) Sc$_{0.03}$Al$_{0.97}$N/GaN, (c)\,Sc$_{0.09}$Al$_{0.91}$N/GaN and (d) Sc$_{0.13}$Al$_{0.87}$N/GaN layers with thicknesses of about 100--120\,nm. The intensity scale of (d) is different from (a)--(c) due to the much lower intensity (i.e., lower crystallinity) of the Sc$_{0.13}$Al$_{0.87}$N 10\=15 reflection.}
\end{figure}

\begin{figure}
	\centering
	\includegraphics[width=0.5\linewidth]{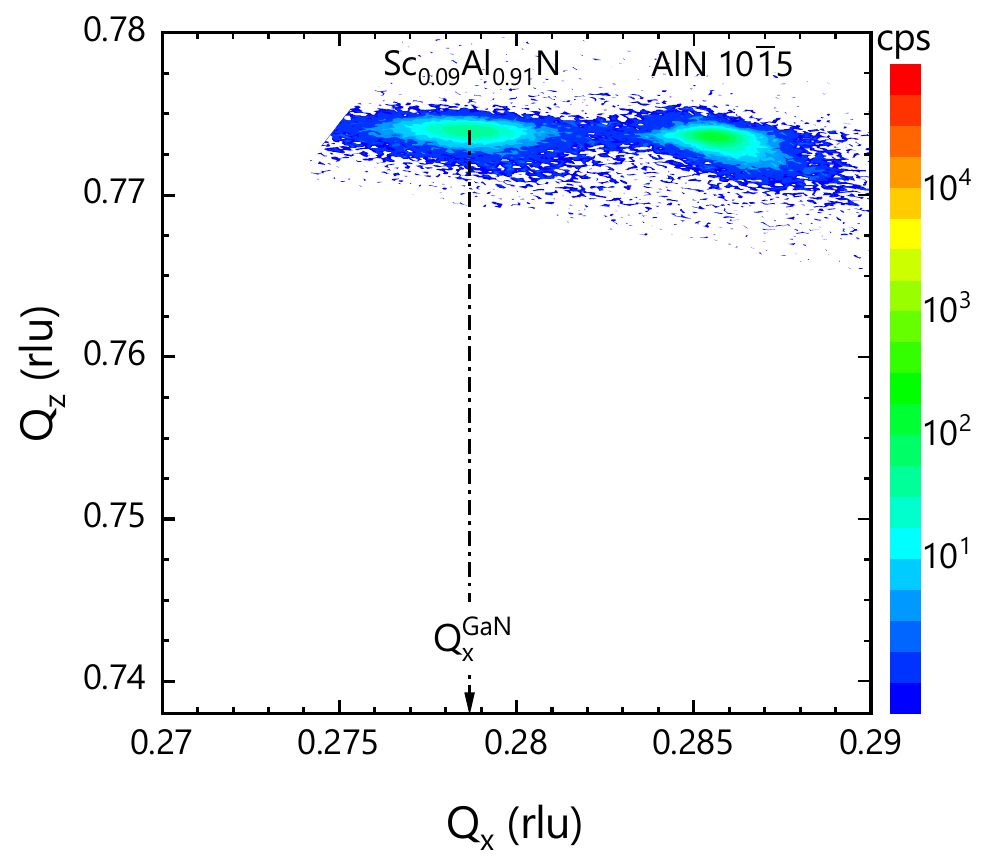}
	\caption{RSM in the vicinity of the asymmetric AlN\,10\=15 reflection of a 350-nm-thick Sc$_{0.09}$Al$_{0.91}$N layer grown on AlN/sapphire(0001) template. The dashed-dotted line indicates the $Q_x$ position of the GaN\,10\=15 reflection, confirming the same \textit{a} lattice constant of Sc$_{0.09}$Al$_{0.91}$N and GaN.}
\end{figure}